\documentclass[sigconf,natbib=true]{acmart}

\author{Bethel Hall\textsuperscript{1}}
\email{bhall2@stevens.edu}
\affiliation{%
  \institution{Stevens Institute of Technology}
  \city{Hoboken}
  \state{New Jersey}
  \country{USA}
}

\author{Sachi Shome\textsuperscript{2}}
\email{sshome@stevens.edu}
\affiliation{%
  \institution{Stevens Institute of Technology}
  \city{Hoboken}
  \state{New Jersey}
  \country{USA}
}

\author{William Eiers\textsuperscript{3}}
\email{weiers@stevens.edu}
\affiliation{%
  \institution{Stevens Institute of Technology}
  \city{Hoboken}
  \state{New Jersey}
  \country{USA}
}

\renewcommand{\shortauthors}{Hall et al.}
\AtBeginDocument{%
  }
\usepackage{listings}
\usepackage{xcolor}
\usepackage{pifont}
\usepackage{bm}
\usepackage{booktabs}
\usepackage{array}
\usepackage[table]{xcolor}
\usepackage{siunitx}
\usepackage{graphicx}
\usepackage{multirow}
\definecolor{vqkey}{RGB}{30,64,175}      
\definecolor{vqok}{RGB}{21,128,61}       
\definecolor{vqbad}{RGB}{185,28,28}      
\definecolor{vqgray}{RGB}{110,110,110}
\definecolor{gridln}{RGB}{200,200,200} 
\usepackage{booktabs}
\usepackage[table]{xcolor}
\usepackage{siunitx}
\usepackage{tabularx}
\usepackage{hhline}
\definecolor{ourshl}{HTML}{EAF0FF}   
\definecolor{facol}{HTML}{FBECEC}    
\definecolor{gainhl}{HTML}{E8F5E9}   
\usepackage[table]{xcolor}
\definecolor{stmtblue}{HTML}{1D4ED8}
\definecolor{stmtteal}{HTML}{0F766E}
\definecolor{stmtred}{HTML}{B91C1C}
\usepackage{booktabs}
\usepackage[table]{xcolor}
\usepackage{siunitx}
\usepackage{subcaption}
\definecolor{ourshl}{HTML}{EEF3FF}   

\newcommand{\blk}[1]{\textbf{#1}}

\usepackage{tabularx}
\usepackage{array}
\usepackage{hhline}
\usepackage[T1]{fontenc}
\usepackage{booktabs}
\usepackage{multirow}
\usepackage{array}
\usepackage[table]{xcolor}
\usepackage{graphicx}
\usepackage{caption}
\usepackage{tgheros}

\usepackage{hyperref}
\usepackage{fontawesome5}
\usepackage{graphicx}

\usepackage{seqsplit}

\usepackage{graphicx}

\begin{document}

\title[VeriFin: Verifying LLM-Generated Financial Claims]{%
  VeriFin: A Neurosymbolic Framework for Verifying
  LLM-Generated Financial Claims%
  \\[0.4em]
  {\large\normalfont
    \href{https://huggingface.co/datasets/bethelhall/XBRLFiling}{%
      \raisebox{-0.2em}{%
        \includegraphics[height=1.2em]{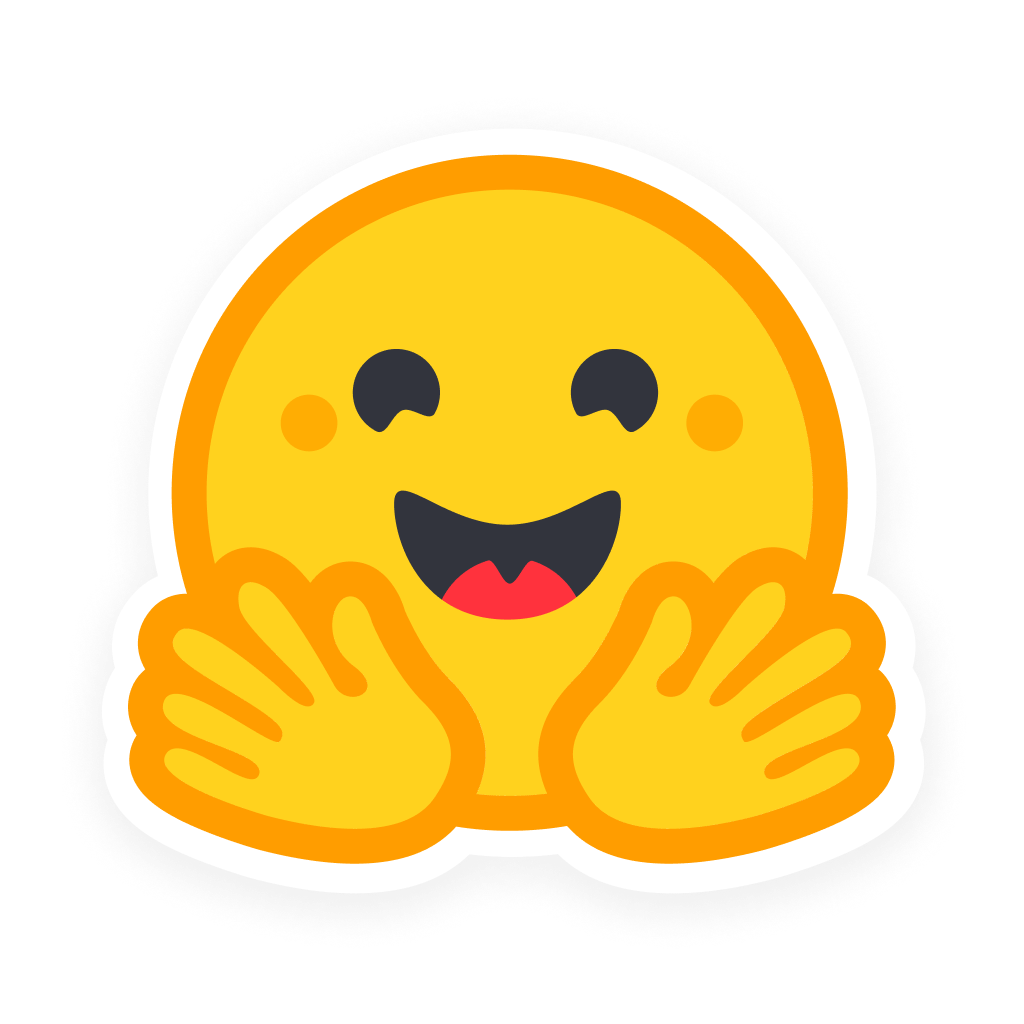}%
      }%
      \hspace{0.3em}XBRLFiling%
    }%
  }%
}

\begin{abstract}
Large Language Models (LLMs) are being adopted in
financial workflows, where accuracy and reliability
are critical. One such application is financial question answering over public company filings. However, LLMs struggle with
quantitative financial reasoning, including identifying financial facts,
interpreting reporting periods and units, and determining
appropriate metrics, leading to incorrect answers.
Recent RAG-based methods have improved financial question
answering but do not ensure factual and numerical correctness.
LLM-based verifiers may also accept incorrect numerical claims.
To address this, we introduce \textsc{VeriFin}, a neurosymbolic
framework for verifying LLM generated numerical financial claims.
Given a question and generated answer, \textsc{VeriFin}
retrieves filing evidence, grounds values in XBRL facts,
and establishes the required formula independently of the answer.
It then uses the Z3 SMT solver to check the resulting
arithmetic constraints. \textsc{VeriFin} accepts consistent
claims, rejects inconsistent claims, and abstains when
required facts or formulas cannot be established.
To evaluate \textsc{VeriFin}, we constructed \textsc{XBRLFiling},
a dataset of 600 numerical questions from corporate filings,
and selected 67 and 403 questions from \textsc{FinanceBench}
and \textsc{TAT-DQA}, respectively. Across six answer models
and three datasets, \textsc{VeriFin} achieves 100\% observed
precision among accepted answers with known correctness.
For Qwen3-30B on \textsc{XBRLFiling}, \textsc{VeriFin}
accepts 304 correct answers versus 166 for an LLM judge,
while reducing false accepts from 87 to zero.
We release our benchmark to support research on reliable
financial question answering.

\end{abstract}


\maketitle
\section{Introduction}

Financial professionals increasingly use large language models to answer numerical questions over public company filings. These questions often require identifying the correct line items, reporting period, unit, and computation from 10-K and 10-Q disclosures. However, a fluent answer may rely on the wrong year, scale, sign, metric definition, or source fact while remaining entirely plausible. Once incorporated into a spreadsheet, valuation model, risk report, compliance assessment, or investment memo, such an error may be difficult to detect and trace. The central reliability problem is therefore not only whether an LLM can generate a financial claim, but whether that claim should be accepted into a financial workflow or used in financial decision-making. For practical use, a plausible numerical answer is not enough; it must be independently verified before acceptance.

Recent works have improved financial question answering through retrieval, structured XBRL access, calculation tools, and executable
programs
\citep{islam2023financebench,han2024xbrl,lai2025sec,wang2025finsage}, nonetheless, numerical reliability remains a challenge. FinanceBench,
XBRL-Agent, and Fin-RATE show that errors continue even when models
have access to filings and external tools, particularly when questions
require the correct entity, reporting period, line item, or computation
\citep{islam2023financebench,han2024xbrl,jiang2026fin}. The problem is therefore not only retrieving the relevant evidence or executing
arithmetic, but establishing whether the selected facts, units, periods,
and formula actually supports a candidate answer. Complementary work on
atomic claim grounding, financial-statement consistency, and
XBRL-based audit verification addresses parts of this problem
\citep{guo2026finground,panda2026finverbench,wang2026auditflow};
however, a model-generated program may execute the wrong computation,
retrieved evidence may be insufficient, and an LLM judge may still
approve a claim that does not follow from the filing. We call the
acceptance of such an incorrect numerical claim a \emph{false accept}.
This motivates the operational question at the center of our work:
\textbf{given a fixed numerical claim produced by an LLM, should a
financial workflow accept it?}


\begin{figure}[t]
  \centering
  \includegraphics[width=\columnwidth]{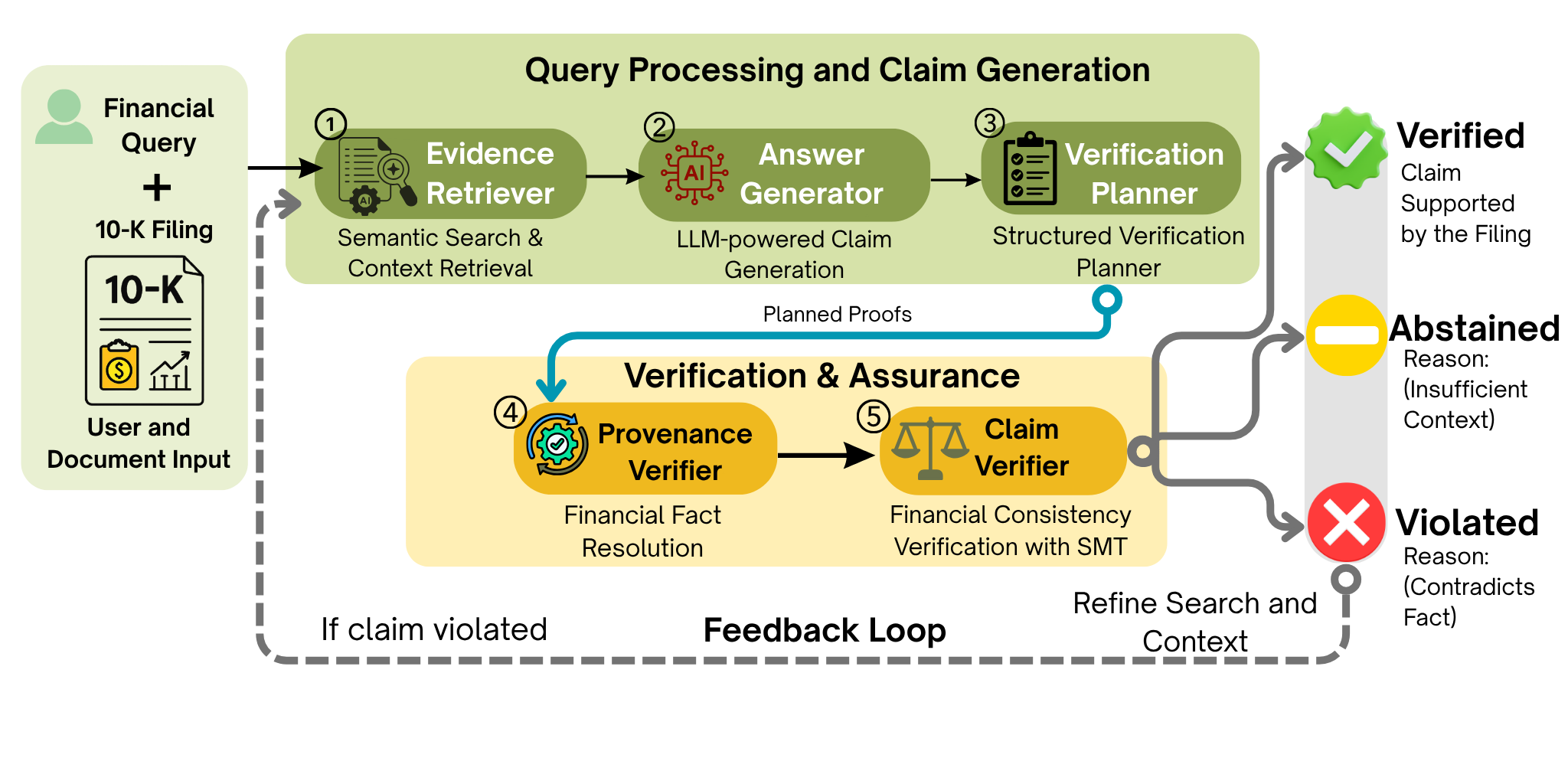}
  \caption{Overview of \textsc{VeriFin}}
  \Description{Overview diagram of the VeriFin framework.}
  \label{fig:overview}
\end{figure}

To address this question, we introduce \textsc{VeriFin}, a filing-grounded neurosymbolic framework that separates claim generation from verification and determines whether numerical claims produced by LLMs should be accepted (Figure~\ref{fig:overview}).
\textsc{VeriFin} treats each candidate claim as an untrusted claim.
First, it grounds the required financial values in filed XBRL facts,
preserving their concepts, reporting periods, units, and contexts.
Second, it establishes the required calculation independently of the
candidate claim. The calculation is obtained from an operation stated
in the question, the filing's XBRL calculation relationships, a
documented financial metric definition, or a filing-specific
definition. Third, \textsc{VeriFin} compiles the grounded values,
calculation, and candidate claim into arithmetic constraints checked
by the Satisfiability Modulo Theories (SMT) solver Z3~\citep{de2008z3}.
The framework returns \textsc{Verified} when the candidate is
consistent with the established facts and calculation,
\textsc{Violated} when it is inconsistent, and \textsc{Abstain} when
the necessary facts or a source-backed calculation cannot be
established unambiguously. Once the facts and calculation are fixed,
the solver decision is deterministic and traceable to the filing
evidence. Abstention is therefore an explicit financial-control
decision: when support for a number cannot be established,
\textsc{VeriFin} declines to approve it rather than substituting a
model-generated assumption.

We evaluate \textsc{VeriFin} on \textsc{XBRLFiling}, a new benchmark of
600 numerical questions constructed from calculations in 10-K filings
of 28 U.S.\ companies, and on a numerical subset of FinanceBench.
On \textsc{XBRLFiling}, \textsc{VeriFin} records zero observed false
accepts, whereas the four baselines accept between 6 and 92 of the
same 92 incorrect claims. \textsc{VeriFin} achieves $92.2\%$ accuracy
among decided cases with $1.2\%$ abstention. On FinanceBench,
\textsc{VeriFin} again records zero observed false accepts, whereas
the baselines accept between 4 and 21 of the same 21 incorrect claims.
The zero false-accept result persists across multiple answer-generator
models. These findings shift the reliability question in LLM-based
financial analysis from whether a claim sounds plausible to whether
filed facts and an independently established calculation support its
acceptance.
We evaluate \textsc{VeriFin} across six answer models on
\textsc{XBRLFiling}, our new benchmark of 600 questions,
and selected numerical subsets of
\textsc{FinanceBench}~\citep{islam2023financebench}
and \textsc{TAT-DQA}~\citep{zhu2022towards}.
Across all 18 experimental settings, we find
100\% precision among accepted answers with known
correctness. Our strongest comparison shows improvements
in both precision and the acceptance of correct answers:
for Qwen3-30B on \textsc{XBRLFiling}, \textsc{VeriFin}
accepts 304 correct answers compared with 166 for LLM Judge,
while reducing false accepts from 87 to zero.
This represents 83\% more correct answers accepted from
the same candidate pool. These results demonstrate that
effective verification can improve the reliability of
accepted answers while allowing more correct answers
to pass.

Our contributions are: 
\begin{itemize}
   \item  We formulate verification of numerical financial claims as an
    acceptance-control problem, focused on the risk of accepting
    incorrect claims and the need to abstain when evidence is
    insufficient. The objective is to reduce the risk of passing
    unsupported claims into downstream financial workflows.

    \item We develop \textsc{VeriFin}, a framework that grounds required values
    in filed XBRL facts, establishes the financial calculation
    independently of the candidate claim, and checks the resulting
    claim using a symbolic solver.
    \item We introduce \textsc{XBRLFiling}, a 600-question benchmark
    constructed from corporate filings, and evaluate \textsc{VeriFin}
    on \textsc{XBRLFiling}, \textsc{FinanceBench}, and \textsc{TAT-DQA}.
    \end{itemize}

\begin{figure*}
  \centering
  \includegraphics[
    width=\textwidth,
    height=0.40\textheight,
    keepaspectratio
  ]{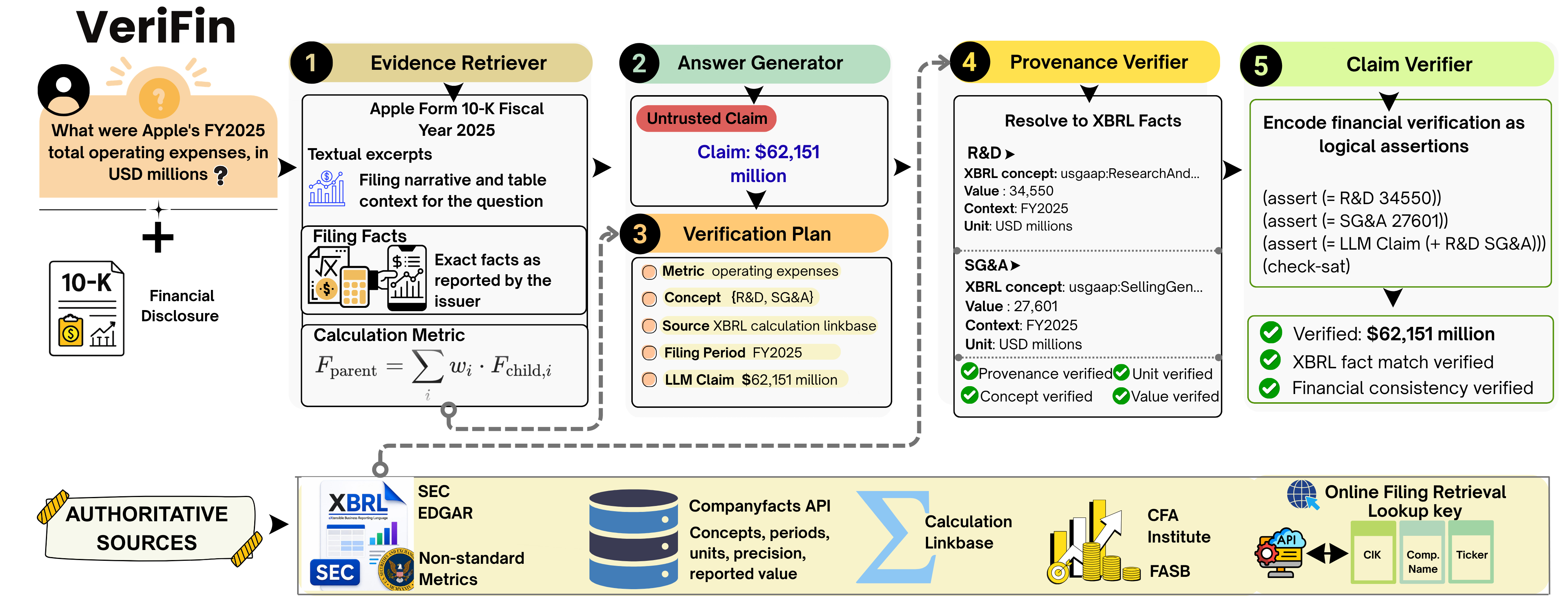}
  \caption{\textsc{VeriFin} applies formal verification to numerical
claims generated by LLMs. For an Apple FY2025 operating-expense query,
it retrieves filing evidence and obtains an untrusted candidate claim.
A Verification Planner specifies the metric, period, unit, operands,
and computation. The Provenance Verifier independently grounds the
operands in XBRL facts and authorizes the formula from
candidate-independent sources. The Claim Verifier then checks the
resulting constraints and returns \textsc{Verified},
\textsc{Violated}, or \textsc{Abstain} with traceable evidence.}
  \label{fig:teaser}
  \Description{The \textsc{VeriFin} pipeline diagram showing the
  process of verifying LLM claims using XBRL facts and an SMT solver.}
\end{figure*}

\paragraph{Neurosymbolic Verification.}
Logic-LM, LINC, and SatLM translate natural-language problems into
formal representations and delegate inference to symbolic solvers
\citep{pan2023logic,olausson2023linc,ye2023satlm}. Similarly, ARc uses
inference-time autoformalization to verify natural-language policy rules
with automated reasoning tools~\citep{bayless2025neurosymbolic}. Although inference is
deterministic with respect to the generated formal representation,
semantic-parsing errors can still lead to incorrect conclusions.
\textsc{VeriFin} reduces reliance on unconstrained free-form translation
by grounding numerical values in XBRL facts and restricting computations
to operations specified in the question, filing calculation relationships,
or documented financial definitions. Given a complete and correctly
grounded verification plan, \textsc{VeriFin} uses Z3
\citep{de2008z3} to deterministically check whether the candidate claim
satisfies the resulting arithmetic constraints.

\section{Related Work}
\label{sec:related_work}

\paragraph{Factuality and LLM-based evaluation.}
Benchmarks such as TruthfulQA and FActScore demonstrate that fluent
language-model outputs are not necessarily factually supported
\citep{lin2022truthfulqa,min2023factscore}. LLM-as-a-judge methods use
one language model to evaluate the output of another, providing a
flexible alternative to task-specific evaluators. However, their
decisions can exhibit systematic biases and reasoning failures
\citep{zheng2023judging}. In numerical financial settings, these
limitations are particularly consequential because a plausible
explanation may conceal an incorrect operand, reporting period, unit,
or calculation.

\paragraph{Financial claim verification.}
FinDVer\citep{zhao2024findver} studies explainable claim verification over long,
hybrid-content financial documents under long-context and
retrieval-augmented settings. Its RAG-CoT approach conditions an LLM
on retrieved evidence to produce an explained binary verdict
\citep{zhao2024findver}. VerifierFC\citep{chungkham2025think} addresses reasoning drift in
complex numerical claim verification through test-time scaling,
generating multiple reasoning paths and using a trained path ranker to
select the highest-scoring path and associated verdict
\citep{chungkham2025think}. These approaches rely on model-generated
judgments or reasoning paths, whereas \textsc{VeriFin} checks a fixed
candidate claim against independently established financial evidence
and arithmetic constraints.

\paragraph{Executable and guided numerical reasoning.}
Program-of-Thoughts delegates arithmetic reasoning to executable
model-generated programs \citep{chen2022program}, and SEC-QA applies
this paradigm to numerical questions over SEC filings
\citep{lai2025sec}. These approaches separate numerical execution from
natural-language generation, but the generated program still specifies
the operands and computation to execute. VERAFI incorporates
formalized financial policies as guidance during agentic generation
\citep{akinfaderin2025verafi}. These methods are complementary to
post-generation verification, where the objective is to determine
whether an already generated numerical claim is sufficiently supported
for acceptance.

\section{Methodology}
\label{sec:methodology}

\textsc{VeriFin} separates answer generation from verification.
Given a financial question, a candidate numerical answer, and
filing evidence, it checks whether the answer follows from
supported financial facts and a documented calculation.
The framework accepts consistent claims, rejects inconsistent
claims, and abstains when the required support cannot be
established.

\subsection{Neurosymbolic Verification}
\label{sec:paradigm}

Financial verification requires identifying the appropriate
concepts, reporting periods, units, and calculation from the filings, then
checking whether they support the generated claim.
As illustrated in Figure~\ref{fig:overview}, we use language models to
semantically interpret the question and propose verification plans.
Then, these proposals are checked against filing evidence,
XBRL calculation relationships and documented metric
definitions to establish the financial inputs and calculations
independently of the candidate claim.

The supported facts, calculation, and candidate value are
encoded as constraints and checked for consistency using
the SMT solver Z3. The decision of the solver is conditional 
on the facts and calculation encoded in those constraints.
We next describe the five components of the framework.

\subsection{Evidence Retriever}
Given a query $q$ and filing document $D$, the Evidence Retriever selects a bounded set of $K$ evidence chunks $E = \{e_1, \dots, e_K\} \subseteq D$ to ground claim generation and verification. To prevent cross-filing contamination, candidates are strictly restricted to $D$ by document identifier. Rather than searching with $q$ alone, the retriever identifies required line-item facts $R = \{r_1, \dots, r_m\}$ and constructs expanded queries $q_i = [q \,||\, r_i \,||\, \text{period} \,||\, \text{statement}]$. Chunks $c \in D$ are scored using a hybrid function:
\begin{equation}
S(q_i, c) = \cos\left(\mathbf{e}(q_i), \mathbf{e}(c)\right) + \lambda \cdot S_{\text{lex}}(q_i, c)
\end{equation}
combining dense cosine similarity with a lexical score $S_{\text{lex}}$ that rewards exact label matches in table rows. Top candidate chunks are reranked via a cross-encoder model. The selected passages and table rows are concatenated to form the
evidence text supplied to the subsequent claim-generation and
verification stages.

\subsection{Answer Generator}
Given the retrieved evidence context $E$, a language model drafts a candidate claim $a$ to query $q$. This stage is intentionally decoupled from decision-making: $a$ is treated as an untrusted claim. To prevent circular evaluation, candidate $a$ is passed forward strictly as a target for formal verification and exerts no influence on the authoritative facts, linkbases, or solver constraints used to check it.

\subsection{Verification Planner}
\label{sec:planner}
The Verification Planner uses an LLM to parse question $q$ and evidence $E$ into a candidate verification plan $C = (M, F, R, \bm{u}, \bm{\rho}, S)$. Defining target metrics $M$, authorized formulas $F$, operand roles $R$, expected units $\bm{u}$, reporting periods $\bm{\rho}$, and formula sources $S$ as sets enables a single plan to naturally support multi-metric and multi-step computational queries.

Formula authority follows a strict hierarchy over non-generative sources (\S\ref{sec:provenance}): explicit operations in $q$, the filing's calculation linkbase, the curated metric registry, or disclosure text. If any formula in $F$ lacks complete and unambiguous authority, \textsc{VeriFin} abstains.

\subsection{Provenance Verifier}\label{sec:provenance}

\paragraph{Grounding Operands in XBRL Facts.}
For each required operand role $r_i$, the Provenance Verifier
independently establishes four properties: the referenced taxonomy
concept, the reporting period, the reported unit, and the filed value.

\emph{Concept.} The role is resolved to a taxonomy concept in three
stages. If the filing's calculation linkbase identifies a child concept
for the role under the requested subtotal, that concept is bound
directly. Otherwise the role is matched against a curated concept
registry that maps normalized financial concepts to candidate taxonomy
elements (e.g., \texttt{us-gaap:AccountsPayableCurrent}). If neither
stage resolves the role, a symbolic label-matching heuristic tokenizes
the filing row label and candidate concept labels, accepts a concept
only when the row's tokens are a subset of the concept label, and
breaks ties by selecting the candidate with the fewest surplus tokens.

\emph{Period and unit.} A concept alone does not uniquely identify a
fact: the same concept is reported for multiple fiscal periods and may
be presented at different display scales (e.g., thousands or millions
of USD). A role therefore binds only to a fact whose reporting period
matches the question and whose unit matches the expected measure, after
normalizing the reported value to the requested scale. A concept that
exists in the filing but has no fact at the required period and unit
does not bind.

\emph{Value and provenance.} The operand value is taken exclusively
from the filed fact. Registry-resolved
concepts are therefore bound \emph{value-blind}: the filed value
overrides whatever figure the model extracted. Consequently,
structural mismatches (e.g., binding \texttt{\seqsplit{ProfitLoss}} where the
authorized policy requires \texttt{NetIncomeLoss}) are not filtered
during grounding but instead surface later as solver constraint
violations. Only the heuristic third stage is guarded by a plausibility
check against the extracted value. Every binding records the concept,
reporting period, unit, accession number, and source fact identifier,
so each admitted operand is fully traceable to a specific reported
filing fact. If any required role cannot be grounded, no verification
plan is admitted and \textsc{VeriFin} abstains.

\paragraph{Establishing Formula Provenance.}
For linkbase-derived subtotals, issuer-declared child concepts and arc
weights $w_i \in \{-1,+1\}$ define the formula directly as
$f(x_1,\dots,x_n)=\sum_{i=1}^{n} w_i x_i$.
Where no linkbase relationship applies, formula authority is derived
from an explicit question template, a curated metric registry,\footnote{
The registry compiles derived financial-metric definitions from CFA Institute curriculum materials.}
or an explicit definition in the filing text. If no authority yields a
complete and unambiguous formula, \textsc{VeriFin} abstains.

Admitted plans record the formula provenance, source identifier, authorized equation $f$, and grounded operand roles, providing downstream verification with an independently established computational baseline.

\subsection{Claim Verifier}
\label{sec:verifier}
Given a validated verification plan and grounded XBRL facts, the Claim
Verifier binds each operand $x_i$ to its filed value $m_i^\star$, computes the
expected value $\hat{v}$ using formula $f$, and compares it with the candidate
value $v_c$:

\[
\Phi =
\bigwedge_i(x_i=m_i^\star)
\wedge(\hat{v}=f(x))
\wedge(|\hat{v}-v_c|\leq\tau),
\]
where
 $\tau$ is the reporting precision. The constraints are compiled into quantifier-free
real arithmetic and evaluated by Z3~\citep{de2008z3}.
The tolerance $\tau$ is half a unit in the last place reported by the
filing, so it admits rounding at the disclosed precision. For a figure stated to the nearest million, $\tau = 0.5$
million; for a percentage stated to one decimal, $\tau = 0.05$
percentage points.

A satisfiable constraint system yields \textsc{Verified}; an unsatisfiable constraint
system yields \textsc{Violated}; the system returns \textsc{Abstain} when
a required operand or authorized formula cannot be established or when the
solver returns \textsc{unknown}. Constraints are named so that a violation can
be accompanied by an unsatisfiable core identifying conflicting filing,
formula, or claim constraints. The final solver decision is deterministic
conditional on the grounded facts and authorized computation.

\paragraph{\textbf{Verifier feedback}.}
Each constraint is named so that a failed verification can be traced to the
relevant filing value, formula condition, or claim bound. For a
verified claim, the Z3 SMT Solver returns logically grounded
evidence supporting the result. For a \textsc{Violated} claim, it identifies
a conflicting subset of named filing, formula, and claim constraints that cannot all
hold together, thereby pinpointing the source of the error. The violated claim
bound further indicates whether the claim overstates or understates the
correct value. For \textsc{Abstain}, the verifier provides the exact requirement it could not
establish, such as a missing formula or facts that cannot be grounded in
the filing. This provides the language model with a concrete target for repair.

\definecolor{XBRLInk}{HTML}{2C4655}
\definecolor{XBRLRule}{HTML}{AEBBC3}
\definecolor{XBRLHeader}{HTML}{EDF2F5}
\definecolor{XBRLInput}{HTML}{F4F7F8}
\definecolor{XBRLTarget}{HTML}{E7EFF3}
\definecolor{XBRLclaim}{HTML}{EDF4EF}
\definecolor{XBRLclaimRule}{HTML}{9CAF9F}

\sisetup{
  group-separator = {,},
  group-minimum-digits = 4,
  detect-weight = true,
  detect-family = true,
  bracket-negative-numbers = true
}

\newcolumntype{A}{S[
  table-format = 6.0,
  table-number-alignment = right,
  table-space-text-pre = {(},
  table-space-text-post = {)}
]}

\newcommand{\xbrlitem}[1]{%
  \hangindent=0.7em\hangafter=1\hspace*{0.7em}#1%
}

\begin{figure}[t]
\centering
\begin{minipage}{\columnwidth}
\setlength{\parindent}{0pt}

{\small\textbf{\textcolor{XBRLInk}{XBRLFiling Example}}}\\[-2pt]
    {\scriptsize\textit{Income statement}}

\vspace{2pt}
\rule{\linewidth}{0.5pt}

\vspace{3pt}

\vspace{3pt}
\textcolor{XBRLInk}{\textbf{Question.}}
What were Apple's FY2025 total operating expenses?

\vspace{2pt}
\vspace{6pt}
{\centering
\footnotesize\textbf{APPLE INC.}\\[-1pt]
\scriptsize\textbf{CONSOLIDATED STATEMENTS OF OPERATIONS}\\[-1pt]
\scriptsize\textit{(USD millions)}\par
}

\vspace{3pt}

\begingroup
\fontsize{7.2}{8.4}\selectfont
\setlength{\tabcolsep}{2pt}
\renewcommand{\arraystretch}{1.05}
\arrayrulecolor{XBRLRule}

\begin{tabularx}{\linewidth}{
@{}>{\raggedright\arraybackslash}X A A A@{}}

\rowcolor{XBRLHeader}
& \multicolumn{3}{c}{\textit{Fiscal years ended}} \\

\rowcolor{XBRLHeader}
& \textbf{2025} & \textbf{2024} & \textbf{2023} \\
\hline

Revenue
& 416161 & 391035 & 383285 \\

Cost of goods sold
& 220960 & 210352 & 214137 \\
\cline{2-4}

\textbf{Gross profit}
& \bfseries 195201 & \bfseries 180683 & \bfseries 169148 \\

\rowcolor{XBRLHeader}
\multicolumn{4}{@{}l}{\textit{Operating expenses:}} \\

\rowcolor{XBRLInput}
\xbrlitem{Research and development}
& \bfseries 34550 & 31370 & 29915 \\

\rowcolor{XBRLInput}
\xbrlitem{Selling, general and administrative}
& \bfseries 27601 & 26097 & 24932 \\
\cline{2-4}

\rowcolor{XBRLTarget}
\xbrlitem{\textbf{Total operating expenses}}
& \multicolumn{1}{c}{\textbf{\textit{[redacted]}}}
& \bfseries 57467
& \bfseries 54847 \\
\cline{2-4}

\textbf{Operating income}
& \bfseries 133050 & \bfseries 123216 & \bfseries 114301 \\

Nonoperating income (expense)
& \textnormal{321}
& 269
& \textnormal{565} \\

Income before taxes
& 132729 & 123485 & 113736 \\

Income tax expense
& 20719 & 29749 & 16741 \\
\cline{2-4}

\textbf{Net income}
& \bfseries 112010 & \bfseries 93736 & \bfseries 96995 \\
\hhline{~===}

\end{tabularx}
\endgroup

\vspace{5pt}
\fcolorbox{XBRLclaimRule}{XBRLclaim}{
\parbox{\dimexpr\linewidth-2\fboxsep-2\fboxrule}{
\footnotesize
\textbf{\textcolor{XBRLInk}{gold answer:}}
$\;34{,}550 + 27{,}601 = \mathbf{62{,}151}$ million
}
}

\end{minipage}

\caption{An XBRLFiling income-statement numerical question: the
target subtotal is redacted, and the complete filing table is
reorganized into a compact listing while preserving all line items and
reporting periods. }

\label{fig:xbrlfiling-example}
\end{figure}

\section{Experimental Setup}

\label{sec:experiments}
We evaluate \textsc{VeriFin} across six answer language models
and three datasets. Baseline comparisons cover four
of these models. Within each comparison, all methods
evaluate the same questions and fixed candidate answers,
without regenerating the candidates. We describe each
method's retrieval procedure and model configuration below.

\subsection{Datasets}

\paragraph{XBRLFiling.} We introduce \textsc{XBRLFiling}, a benchmark of 600 questions
constructed from calculation relationships declared in corporate
filings. Each question includes financial evidence, a reference
answer, and the corresponding calculation. We reduct the required facts in the supplied evidence while retaining the evidence,
requiring the answer to be computed rather than copied. The benchmark covers 15 financial metrics across income statements,
balance sheets, and cash-flow statements (see Table~\ref{tab:topics}).

\begin{figure*}[t]
  \centering

  \begin{subfigure}[t]{0.57\textwidth}
    \centering
    \includegraphics[width=\linewidth]
      {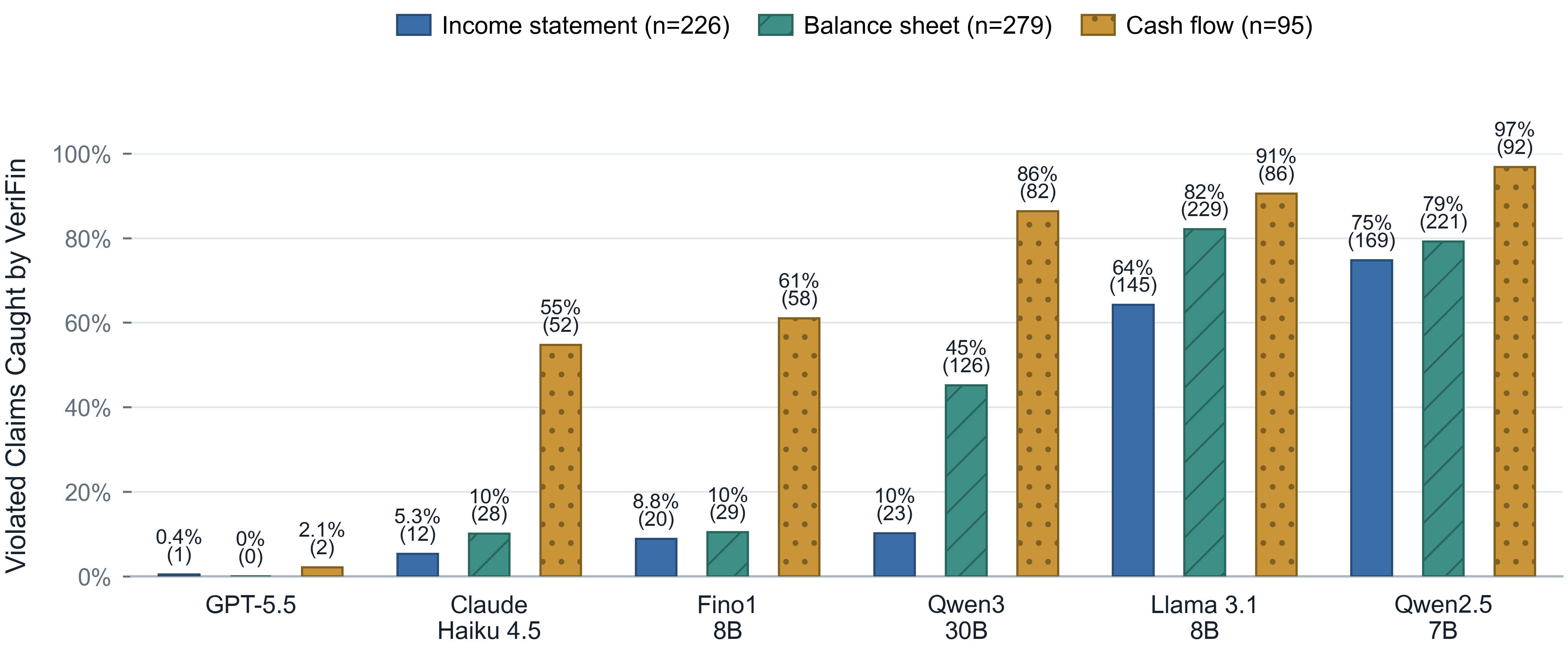}
    \label{fig:xbrl-error-profile}
  \end{subfigure}%
  \hspace{0.02\textwidth}%
  \begin{subfigure}[t]{0.39\textwidth}
    \centering
    \includegraphics[width=\linewidth]
      {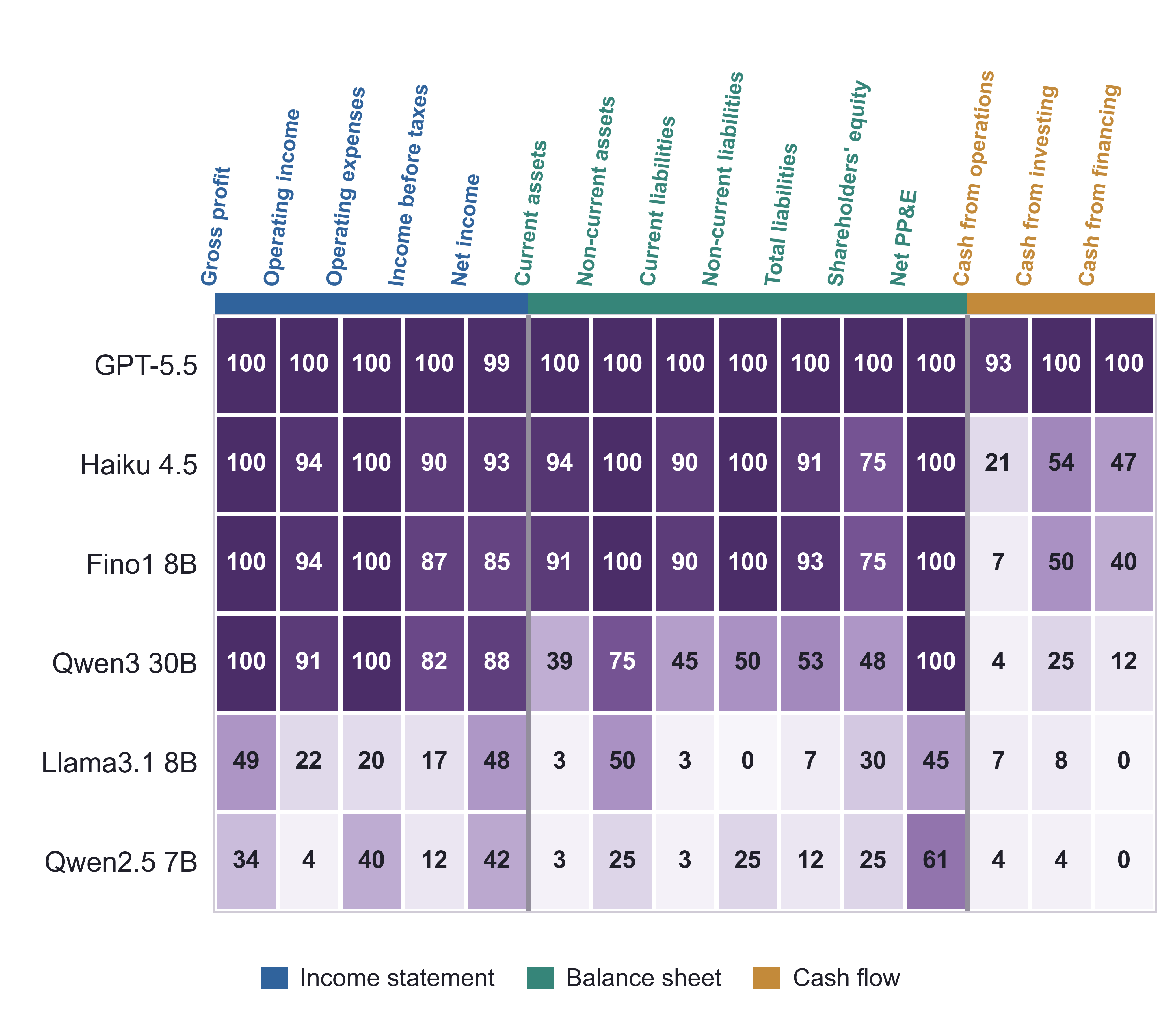}
    \label{fig:test}
  \end{subfigure}

  \caption{Left: Percentage of incorrect claims generated by each LLM and rejected by \textsc{VeriFin} on \textsc{XBRLFiling.} Right: Candidate-answer accuracy by requested line item on \textsc{XBRLFiling}. Generator errors are strongly metric-dependent. Cash-flow subtotals remain difficult for most evaluated models, while every incorrect candidate represented in the left panel is rejected by \textsc{VeriFin}.}
  \label{fig:combined-performance}
\end{figure*}

\definecolor{ourshl}{RGB}{234,242,248}  
\definecolor{facol}{RGB}{246,239,233}   

\definecolor{gainhl}{HTML}{E8F5E9}   

\paragraph{FinanceBench and TAT-DQA}
We additionally evaluate on selected numerical subsets of
\textsc{FinanceBench}~\citep{islam2023financebench}
and \textsc{TAT-DQA}~\citep{zhu2022towards},
containing 67 and 403 questions, respectively.
TAT-DQA is a financial document question-answering
benchmark combining semi-structured tables and
unstructured text from financial reports.
We select questions requiring financial value extraction,
numerical comparisons, or calculations.
We exclude questions requiring only qualitative
interpretation because our framework targets numerical
financial claims that remain challenging for
LLMs~\citep{han2024xbrl}.
For \textsc{TAT-DQA}, we further require matching SEC
filings with available XBRL evidence and apply
eligibility checks before running verification.
These selected subsets extend our evaluation beyond
questions constructed from XBRL calculation relationships.

\begin{table}[t]
\centering
\footnotesize
\setlength{\tabcolsep}{3pt}
\renewcommand{\arraystretch}{1.08}
\begin{tabular}{@{}lrrp{4.6cm}@{}}
\toprule
\textbf{Statement} & \textbf{M} & \textbf{N} & \textbf{Metrics} \\
\midrule
Income
& 5 & 226
& Net, pretax, and operating income; gross profit; operating expenses \\
Balance
& 7 & 279
& Current/non-current assets and liabilities; total liabilities;
shareholders' equity; net PP\&E \\
Cash flow
& 3 & 95
& Operating, investing, and financing cash flows \\
\bottomrule
\end{tabular}
\caption{XBRLFiling composition ($N{=}600$; 15 metrics). M denotes the
number of metrics.}
\label{tab:topics}
\end{table}

\subsection{Evaluated Models and Baselines}
\label{sec:baselines}

\paragraph{Financial Concepts and Formula Registry.}
Financial questions and filings can use different terminology
for the same concept, while derived metrics require calculation
definitions that may not be stated in the filing. We curate a
registry to connect financial terms to filing concepts and
provide documented formulas for verification. This supplies a
reference for checking the concepts and calculations proposed
by the verification planner.

The registry contains 59 metric entries, 17 arithmetic operation
rules, and 126 financial concepts. Entries specify formulas,
required inputs, and supporting references drawn from CFA
Institute materials, US GAAP and IFRS taxonomies, and FinanceBench
question definitions. Nineteen metric entries are specific to
FinanceBench. The registry supplies definitions and mappings;
numerical values are retrieved from the evidence for each question.

\paragraph{Answer Models.}
To evaluate verification robustness across diverse model capabilities, candidate claims are generated using six language models spanning different scales and architectures: frontier closed models (\texttt{GPT-5.5}, \texttt{Claude Haiku-4.5}), open-weight reasoning models (\texttt{Qwen3-30B}, \texttt{Llama-3.1-8B}, \texttt{Qwen2.5-7B}), and domain-specialized financial models (\texttt{Fin-o1-8B}). Evaluating across this models ensures our results reflect verifier performance rather than specific generator biases. We served the openweight claim generation models with vLLM, calling them through
its OpenAI-compatible endpoint; weights were downloaded from HuggingFace. Claude Haiku~4.5 and GPT-5.5 were called through their
respective APIs.  The local models are accessed through OpenAPI compatible endpoints, and Claude Haiku 4.5 through Anthropic API. generation settings vary across models and runs. For GPT-5.5 on \textsc{FinanceBench} and
\textsc{TAT-DQA}, we use low reasoning effort and an
8192-token output budget, without setting temperature.

\paragraph{Baselines.}
We compare \textsc{VeriFin} with an LLM Judge,
FinDVer~\citep{zhao2024findver}, and
VerifierFC~\citep{chungkham2025think}.
All methods evaluate the same frozen candidate claims, which are reused
without regeneration. Evidence retrieval and input construction follow
each method's respective evaluation pipeline; no baseline receives gold
answers, gold evidence, or verdicts produced by another method.
\begin{itemize}
    \item \textbf{LLM judge:} The model receives the question,
    retrieved evidence, and fixed candidate claim, and predicts whether
    the claim should be accepted as correct.
   \item \textbf{FinDVer:}
We apply FinDVer's released RAG-CoT method using its
text-embedding-3-large top-10 retrieval, original prompts, and binary
evaluation procedure. Following FinDVer's evaluation design across
diverse LLM backbones, we use Claude Haiku 4.5 as the verifier while
preserving the remaining components and settings of the released
pipeline. FinDVer produces a binary \emph{entailed}/\emph{refuted} verdict,
which we map to \emph{accept} or \emph{reject}. Because it returns one of
these two decisions for every evaluated claim, it has no abstentions
and therefore achieves $100\%$ decision coverage.

\item \textbf{VerifierFC:}
We reproduce the GPT-4o-mini Best-of-10 configuration reported for
VerifierFC, using its full-filing BM25 retrieval and MiniLM reranking
stages. We select this generator configuration because it achieves the
highest reported ClaimDecomp macro-F1 among the Best-of-10
configurations evaluated in the original study. Following the reported
protocol, GPT-4o-mini generates ten reasoning-path--verdict pairs for
each claim. A Llama-3.2-3B LoRA path ranker, trained on tuples
comprising a claim, an LLM-generated reasoning path, and its predicted
verdict, scores the candidates and selects the highest-scoring verdict.
We map supporting and refuting verdicts to \emph{accept} and
\emph{reject}, respectively, while a conflicting verdict is treated as
an \emph{abstention}; VerifierFC can therefore have less than $100\%$
coverage.   The authors did not release the trained LoRA
checkpoint, we reproduce the ranker locally using their released
training and development data, implementation, and reported
hyperparameters.
\end{itemize}


\paragraph{Inference and scoring.}
LLM Judge and FinDVer predict verification verdicts from
the candidate claim and evidence. VerifierFC uses a trained
verifier to score multiple generated reasoning paths and
returns the verdict associated with the highest-scoring path.
Within each comparison, we evaluate all methods against
the same candidate correctness labels, using the scoring
policy for each dataset. These labels are used only for
evaluation and are not supplied to the verification methods.
The numerical tolerance used by \textsc{VeriFin} during
verification is defined in \S\ref{sec:verifier}.

\paragraph{Metrics.}
True accepts (TA) and false accepts (FA) count correct and
incorrect accepted claims, respectively. True rejects (TR)
and false rejects (FR) count incorrect and correct rejected
claims. Abs counts abstentions, including cases where
candidate generation fails.

We report false accepts, precision
$\mathrm{TA}/(\mathrm{TA}+\mathrm{FA})$, and decision accuracy
$(\mathrm{TA}+\mathrm{TR})/
(\mathrm{TA}+\mathrm{FA}+\mathrm{TR}+\mathrm{FR})$.
Precision measures correctness among accepted claims, while
decision accuracy measures correctness among scored acceptance
and rejection decisions. Coverage is
$(n-\mathrm{Abs})/n$, and the abstention rate is
$\mathrm{Abs}/n$, where $n$ includes every question.
Cases with unresolved correctness are excluded from precision
and decision accuracy but remain in the coverage calculation.

For repair, recovery requires an initially incorrect, rejected
claim to be revised into a correct answer that passes
verification. Naive repair uses one generic retry.
Verifier-guided $\mathrm{pass@1}$ measures recovery on the
first attempt, while $\mathrm{pass@3}$ measures recovery
within three attempts. We report recovery rates together
with recovered/scored counts for each condition. 

\subsection{Exploratory Repair Setup}
\label{sec:repair_setup}

As an auxiliary analysis, we examine whether diagnostics already
produced by \textsc{VeriFin} can support the revision of rejected
numerical claims. When a claim is \textsc{Violated}, the verifier can
use named constraints from the solver's unsatisfiable core to identify
the filed values, arithmetic relation, and claim bound involved in the
inconsistency.

We consider two repair conditions:

\begin{itemize}
    \item \textbf{Naive Repair:} the model receives the question,
    filing evidence, previous answer, and a generic request to revise
    the incorrect calculation.

    \item \textbf{Verifier-Guided Repair:} the model additionally
    receives structured diagnostic feedback derived from the
    verification failure.
\end{itemize}

The naive condition uses one retry. Verifier-guided pass@1 evaluates
the first guided attempt, while pass@3 allows up to three guided
attempts and counts a case as recovered if any attempt succeeds.
Every revised claim is checked using the same verification procedure;
it is counted as recovered only if it matches the gold answer and
receives a \textsc{Verified} verdict.

This exploratory analysis uses archived, condition-specific scored
subsets of true catches rather than a single matched set. We therefore
report the recovery rates in Table~\ref{tab:repair} descriptively and
do not interpret differences between conditions as controlled effects.
\section{Results and Discussion}

\definecolor{gainhl}{HTML}{EDF5EE}
\definecolor{gridln}{HTML}{B8C1C8}

\begin{table}[t]
\centering\footnotesize
\setlength{\tabcolsep}{6pt}
\renewcommand{\arraystretch}{1.08}
\setlength{\aboverulesep}{0pt}
\setlength{\belowrulesep}{0pt}

\begin{tabular}{
  @{}l
  c
  !{\color{gridln}\vrule width 0.4pt}
  cc@{}
}
\toprule
&
\textbf{Naive Repair}
&
\multicolumn{2}{c}{\textbf{Verifier-Guided Repair}}
\\
\cmidrule(lr){3-4}

\textbf{Backbone}
&
\textbf{Recovery (\%)}
&
\textbf{pass@1 (\%)}
&
\textbf{pass@3 (\%)}
\\
\midrule

\blk{XBRLFiling} \\[1pt]

Qwen3-30B        &  6.5 & 46.3 & 47.7 \\
Fin-o1-8B        & 26.1 & 57.0 & 69.9 \\
Claude Haiku 4.5 & 23.1 & 40.5 & 55.6 \\

\midrule
\blk{FinanceBench} \\[1pt]

Qwen3-30B        & 0.0 & 37.5 & 50.0 \\
Fin-o1-8B        & 0.0 &  0.0 &  9.1 \\
Claude Haiku 4.5 & 0.0 & 10.0 & 20.0 \\

\bottomrule
\end{tabular}

\caption{%
Exploratory repair recovery rates (\%) on archived,
condition-specific scored subsets of true catches.
Naive Repair uses one generic retry.
Verifier-guided pass@1 uses one attempt with solver-derived
diagnostic feedback, whereas pass@3 allows up to three guided attempts.
Because the conditions were evaluated on separately scored archived
subsets rather than a single matched set, the results are descriptive
rather than paired comparisons.
}
\label{tab:repair}
\end{table}

\label{sec:results}

\definecolor{baselineGreenBg}{HTML}{E4F2E7}
\definecolor{baselineRedBg}{HTML}{F4CCCC}

\newcommand{\zerofa}[1]{%
  \cellcolor{baselineGreenBg}#1%
}

\newcommand{\nonzerofa}[1]{%
  \cellcolor{baselineRedBg}#1%
}

\newcommand{\sigp}[1]{%
  \cellcolor{baselineGreenBg}\textbf{\ensuremath{#1}}%
}

\newcommand{\nonsigp}[1]{%
  \cellcolor{baselineRedBg}\ensuremath{#1}%
}

\newcommand{\pnum}[2]{%
  #1{\times}10^{#2}%
}

\begin{table*}[t]
\caption{%
Baselines versus VeriFin (ours).
FA counts incorrect answers accepted;
Acc.\ is accuracy among non-abstaining decisions;
Cov.\ is the percentage receiving a non-abstaining decision.
Each p-value shown for a baseline compares its precision with the
VeriFin result at the end of the same model block and is Holm-adjusted.
LLM Judge and FinDVer have 100\% coverage because their binary protocols
produce an accept/reject decision for every question.
Green indicates zero FA or $p<0.05$; red indicates nonzero FA or
$p\geq0.05$.
}
\label{tab:baseline-three-datasets}

\centering
\color{black}
\scriptsize
\setlength{\tabcolsep}{1.3pt}
\renewcommand{\arraystretch}{1.08}

\begin{tabular*}{\textwidth}{
  @{\extracolsep{\fill}}
  ll
  rcccc
  !{\hspace{2pt}\color{black!22}\vrule width 0.4pt\hspace{2pt}}
  rcccc
  !{\hspace{2pt}\color{black!22}\vrule width 0.4pt\hspace{2pt}}
  rcccc
  @{}
}
\toprule
& &
\multicolumn{5}{c}{FinanceBench ($n=67$)} &
\multicolumn{5}{c}{XBRLFiling ($n=600$)} &
\multicolumn{5}{c}{TAT-DQA ($n=403$)}
\\

\cmidrule(lr){3-7}
\cmidrule(lr){8-12}
\cmidrule(l){13-17}

& &
& & \multicolumn{2}{c}{Precision} & &
& & \multicolumn{2}{c}{Precision} & &
& & \multicolumn{2}{c}{Precision} &
\\

\cmidrule(lr){5-6}
\cmidrule(lr){10-11}
\cmidrule(l){15-16}

Model & Method
& FA & Acc.\ (\%) & \% & $p$ & Cov.\ (\%)
& FA & Acc.\ (\%) & \% & $p$ & Cov.\ (\%)
& FA & Acc.\ (\%) & \% & $p$ & Cov.\ (\%)
\\
\midrule

Qwen2.5-7B
& LLM Judge
& \nonzerofa{22} & 49.3 & 46.3
& \sigp{\pnum{6.5}{-3}} & 100.0
& \nonzerofa{167} & 59.3 & 19.3
& \sigp{\pnum{2.7}{-4}} & 100.0
& \nonzerofa{89} & 56.3 & 54.6
& \sigp{\pnum{1.4}{-7}} & 100.0
\\

& FinDVer
& \nonzerofa{13} & 65.7 & 53.6
& \sigp{\pnum{9.4}{-6}} & 100.0
& \nonzerofa{29} & 85.7 & 63.8
& \sigp{\pnum{9.4}{-6}} & 100.0
& \nonzerofa{21} & 72.7 & 81.9
& \sigp{\pnum{8.4}{-8}} & 100.0
\\

& VerifierFC
& \nonzerofa{3} & 66.7 & 57.1
& \nonsigp{\pnum{8.8}{-2}} & 31.3
& \nonzerofa{24} & 53.4 & 31.4
& \sigp{\pnum{1.9}{-15}} & 9.7
& \nonzerofa{19} & 68.1 & 75.0
& \sigp{\pnum{1.3}{-6}} & 52.9
\\

& \textbf{VeriFin}
& \zerofa{0} & 78.8 & 100.0 &  & 49.3
& \zerofa{0} & 87.6 & 100.0 &  & 98.3
& \zerofa{0} & 71.7 & 100.0 &  & 62.3
\\

\addlinespace[2pt]

Qwen3-30B
& LLM Judge
& \nonzerofa{23} & 50.7 & 57.4
& \sigp{\pnum{1.7}{-3}} & 100.0
& \nonzerofa{87} & 51.8 & 65.6
& \sigp{\pnum{5.1}{-5}} & 100.0
& \nonzerofa{44} & 74.7 & 86.0
& \sigp{\pnum{2.1}{-6}} & 100.0
\\

& FinDVer
& \nonzerofa{15} & 47.8 & 54.5
& \sigp{\pnum{3.7}{-5}} & 100.0
& \nonzerofa{34} & 76.5 & 88.4
& \sigp{\pnum{1.3}{-6}} & 100.0
& \nonzerofa{23} & 85.1 & 92.5
& \sigp{\pnum{1.4}{-4}} & 100.0
\\

& VerifierFC
& \nonzerofa{5} & 56.5 & 58.3
& \sigp{\pnum{1.7}{-2}} & 34.3
& \nonzerofa{27} & 69.8 & 74.0
& \sigp{\pnum{7.0}{-6}} & 28.7
& \nonzerofa{36} & 75.8 & 85.5
& \sigp{\pnum{2.3}{-8}} & 73.9
\\

& \textbf{VeriFin}
& \zerofa{0} & 84.6 & 100.0 &  & 77.6
& \zerofa{0} & 89.5 & 100.0 &  & 99.7
& \zerofa{0} & 77.3 & 100.0 &  & 79.7
\\

\addlinespace[2pt]

Llama-3.1-8B
& LLM Judge
& \nonzerofa{34} & 49.3 & 47.7
& \sigp{\pnum{3.8}{-3}} & 100.0
& \nonzerofa{456} & 23.3 & 21.4
& \sigp{\pnum{4.1}{-5}} & 100.0
& \nonzerofa{234} & 41.7 & 39.2
& \sigp{\pnum{1.1}{-8}} & 100.0
\\

& FinDVer
& \nonzerofa{13} & 59.7 & 51.9
& \sigp{\pnum{2.4}{-5}} & 100.0
& \nonzerofa{55} & 83.5 & 59.3
& \sigp{\pnum{1.3}{-9}} & 100.0
& \nonzerofa{24} & 78.9 & 79.1
& \sigp{\pnum{7.1}{-6}} & 100.0
\\

& VerifierFC
& \nonzerofa{2} & 70.4 & 77.8
& \nonsigp{\pnum{3.3}{-1}} & 40.3
& \nonzerofa{38} & 51.4 & 44.1
& \sigp{\pnum{3.3}{-11}} & 17.8
& \nonzerofa{51} & 66.8 & 56.8
& \sigp{\pnum{1.3}{-21}} & 68.0
\\

& \textbf{VeriFin}
& \zerofa{0} & 69.2 & 100.0 &  & 58.2
& \zerofa{0} & 91.4 & 100.0 &  & 97.3
& \zerofa{0} & 80.7 & 100.0 &  & 68.0
\\

\addlinespace[2pt]

GPT-5.5
& LLM Judge
& \nonzerofa{9} & 86.6 & 86.4
& \sigp{\pnum{1.9}{-2}} & 100.0
& \nonzerofa{1} & 92.7 & 99.8
& \nonsigp{\pnum{9.5}{-1}} & 100.0
& \nonzerofa{31} & 92.1 & 92.3
& \sigp{\pnum{1.1}{-4}} & 100.0
\\

& FinDVer
& \nonzerofa{12} & 61.2 & 74.5
& \sigp{\pnum{5.0}{-4}} & 100.0
& \zerofa{0} & 61.0 & 100.0
& \nonsigp{\pnum{1.0}{0}} & 100.0
& \nonzerofa{30} & 90.6 & 92.2
& \sigp{\pnum{7.1}{-6}} & 100.0
\\

& VerifierFC
& \nonzerofa{11} & 63.0 & 71.1
& \sigp{\pnum{8.6}{-4}} & 68.7
& \zerofa{0} & 95.9 & 100.0
& \nonsigp{\pnum{1.0}{0}} & 40.8
& \nonzerofa{19} & 87.4 & 94.3
& \sigp{\pnum{1.4}{-4}} & 88.8
\\

& \textbf{VeriFin}
& \zerofa{0} & 89.4 & 100.0 &  & 70.1
& \zerofa{0} & 93.8 & 100.0 &  & 100.0
& \zerofa{0} & 91.3 & 100.0 &  & 80.1
\\

\bottomrule
\end{tabular*}
\end{table*}

\definecolor{vfgain}{HTML}{D0E6D5}
\definecolor{vfhead}{HTML}{E3EFE7}
\definecolor{vfaccent}{HTML}{2D5F45}
\definecolor{vfgrid}{HTML}{B8C1C8}
\definecolor{vfrule}{HTML}{2D5F45}
\definecolor{vfmuted}{HTML}{394556}
\newcommand{\vfCI}[3]{%
  \shortstack[c]{#1\\{\fontsize{7}{8}\selectfont\textcolor{vfmuted}{#2--#3}}}%
}

\definecolor{gainhl}{HTML}{E4F0E6}
\definecolor{gridln}{HTML}{B8C1C8}

\begin{table*}[t]
\centering\small
\caption{\textsc{VeriFin} performance across six answer generators
and three datasets. TA/FA/FR/Abs are counts; precision, accuracy,
and coverage are percentages with 95\% exact confidence intervals
in brackets.}
\label{tab:models-three-datasets}
\setlength{\tabcolsep}{3.5pt}
\renewcommand{\arraystretch}{0.9}
\newcommand{\rowpad}{\rule{0pt}{2.6ex}}
\setlength{\aboverulesep}{0pt}
\setlength{\belowrulesep}{0pt}
\sisetup{detect-weight=true, mode=text}
\newcommand{\ci}[2]{\rule[-0.9ex]{0pt}{0pt}\scriptsize\color{black!55}[#1,\,#2]}
\resizebox{\textwidth}{!}{%
\begin{tabular}{@{}l
S[table-format=3.0] >{\columncolor{gainhl}}S[table-format=1.0] S[table-format=2.0] S[table-format=2.0] S[table-format=3.1] S[table-format=2.1] S[table-format=3.1]
!{\color{gridln}\vrule width 0.4pt}
S[table-format=2.0] >{\columncolor{gainhl}}S[table-format=1.0] S[table-format=2.0] S[table-format=2.0] S[table-format=3.1] S[table-format=2.1] S[table-format=2.1]
!{\color{gridln}\vrule width 0.4pt}
S[table-format=3.0] >{\columncolor{gainhl}}S[table-format=1.0] S[table-format=2.0] S[table-format=3.0] S[table-format=3.1] S[table-format=2.1] S[table-format=2.1]@{}}
\toprule
\rowpad& \multicolumn{7}{c}{\textbf{XBRLFiling} ($n{=}600$)} & \multicolumn{7}{c}{\textbf{FinanceBench} ($n{=}67$)} & \multicolumn{7}{c}{\textbf{TADQA} ($n{=}403$)} \\
\cmidrule(lr){2-8}\cmidrule(lr){9-15}\cmidrule(l){16-22}
\rowpad\textbf{Claim model}
& {TA} & {FA$\,\downarrow$} & {FR} & {Abs} & {Prec.} & {Acc.} & {Cov.}
& {TA} & {FA$\,\downarrow$} & {FR} & {Abs} & {Prec.} & {Acc.} & {Cov.}
& {TA} & {FA$\,\downarrow$} & {FR} & {Abs} & {Prec.} & {Acc.} & {Cov.} \\
\midrule
\rowpad Qwen2.5-7B & 35 & 0 & 73 & 10 & 100.0 & 87.6 & 98.3
                 & 10 & 0 & 7 & 34 & 100.0 & 78.8 & 49.3
                 & 62 & 0 & 71 & 152 & 100.0 & 71.7 & 62.3 \\
 &  &  &  &  & {\ci{90.0}{100.0}} & {\ci{84.7}{90.2}} & {\ci{97.0}{99.2}}
                 &  &  &  &  & {\ci{69.2}{100.0}} & {\ci{61.1}{91.0}} & {\ci{36.8}{61.8}}
                 &  &  &  &  & {\ci{94.2}{100.0}} & {\ci{65.7}{77.2}} & {\ci{57.4}{67.0}} \\
\rowpad Llama-3.1-8B & 74 & 0 & 50 & 16 & 100.0 & 91.4 & 97.3
                 & 10 & 0 & 12 & 28 & 100.0 & 69.2 & 58.2
                 & 48 & 0 & 53 & 129 & 100.0 & 80.7 & 68.0 \\
 &  &  &  &  & {\ci{95.1}{100.0}} & {\ci{88.9}{93.6}} & {\ci{95.7}{98.5}}
                 &  &  &  &  & {\ci{69.2}{100.0}} & {\ci{52.4}{83.0}} & {\ci{45.5}{70.2}}
                 &  &  &  &  & {\ci{92.6}{100.0}} & {\ci{75.5}{85.2}} & {\ci{63.2}{72.5}} \\
\rowpad Qwen3-30B & 304 & 0 & 63 & 2 & 100.0 & 89.5 & 99.7
                 & 24 & 0 & 8 & 15 & 100.0 & 84.6 & 77.6
                 & 196 & 0 & 73 & 82 & 100.0 & 77.3 & 79.7 \\
 &  &  &  &  & {\ci{98.8}{100.0}} & {\ci{86.7}{91.8}} & {\ci{98.8}{100.0}}
                 &  &  &  &  & {\ci{85.8}{100.0}} & {\ci{71.9}{93.1}} & {\ci{65.8}{86.9}}
                 &  &  &  &  & {\ci{98.1}{100.0}} & {\ci{72.3}{81.7}} & {\ci{75.4}{83.5}} \\
\rowpad Fin-o1 & 436 & 0 & 49 & 8 & 100.0 & 91.7 & 98.7
                 & 32 & 0 & 9 & 12 & 100.0 & 83.6 & 82.1
                 & 191 & 0 & 61 & 103 & 100.0 & 79.7 & 74.4 \\
 &  &  &  &  & {\ci{99.2}{100.0}} & {\ci{89.2}{93.8}} & {\ci{97.4}{99.4}}
                 &  &  &  &  & {\ci{89.1}{100.0}} & {\ci{71.2}{92.2}} & {\ci{70.8}{90.4}}
                 &  &  &  &  & {\ci{98.1}{100.0}} & {\ci{74.7}{84.1}} & {\ci{69.9}{78.6}} \\
\rowpad Claude Haiku~4.5 & 455 & 0 & 46 & 7 & 100.0 & 92.2 & 98.8
                 & 28 & 0 & 9 & 13 & 100.0 & 83.3 & 80.6
                 & 244 & 0 & 55 & 80 & 100.0 & 83.0 & 80.1 \\
 &  &  &  &  & {\ci{99.2}{100.0}} & {\ci{89.8}{94.3}} & {\ci{97.6}{99.5}}
                 &  &  &  &  & {\ci{87.7}{100.0}} & {\ci{70.7}{92.1}} & {\ci{69.1}{89.2}}
                 &  &  &  &  & {\ci{98.5}{100.0}} & {\ci{78.4}{86.9}} & {\ci{75.9}{83.9}} \\
\rowpad GPT-5.5 & 560 & 0 & 37 & 0 & 100.0 & 93.8 & 100.0
                 & 34 & 0 & 6 & 20 & 100.0 & 86.7 & 70.1
                 & 282 & 0 & 28 & 80 & 100.0 & 91.3 & 80.1 \\
 &  &  &  &  & {\ci{99.3}{100.0}} & {\ci{91.6}{95.6}} & {\ci{99.4}{100.0}}
                 &  &  &  &  & {\ci{89.7}{100.0}} & {\ci{73.2}{94.9}} & {\ci{57.7}{80.7}}
                 &  &  &  &  & {\ci{98.7}{100.0}} & {\ci{87.7}{94.2}} & {\ci{75.9}{83.9}} \\
\bottomrule
\end{tabular}%
}

\label{tab:models-three-datasets}
\end{table*}

\subsection{Comparison with Verification Baselines}

Table~\ref{tab:baseline-three-datasets} shows that
\textsc{VeriFin} is the only evaluated method to record
zero false accepts across all 12 settings covering four
answer models and three datasets. It achieves 100\%
observed precision throughout these comparisons,
including settings where the other methods accept
substantial numbers of incorrect claims.

For answers generated by Qwen3-30B on
\textsc{XBRLFiling}, LLM Judge accepts 87 incorrect
answers, FinDVer accepts 34, and VerifierFC accepts 27,
compared with zero for \textsc{VeriFin}.
At the same time, \textsc{VeriFin} accepts 304 correct
answers compared with 166 for LLM Judge.
This result demonstrates a particularly useful outcome:
the verification pipeline prevents more incorrect
answers from passing while preserving more correct
answers from the same candidate pool.

The benefit also extends to GPT-5.5. Its judge accepts
nine incorrect answers on \textsc{FinanceBench} and
31 on \textsc{TAT-DQA}, whereas \textsc{VeriFin}
accepts none in the corresponding comparisons.
The advantage is smaller on \textsc{XBRLFiling},
where FinDVer, VerifierFC, and \textsc{VeriFin}
all record zero false accepts and LLM Judge records
only one. The precision difference from LLM Judge
is not statistically significant in that setting.
Overall, the results demonstrate consistent control
of observed false accepts across the evaluated settings,
with additional gains in retaining correct answers
in some comparisons.
\subsection{Performance Across Claim Generating Models}

Table~\ref{tab:models-three-datasets} reports
\textsc{VeriFin}'s performance across six answer models
and three datasets. Across all 18 settings, none of
the accepted answers whose correctness could be assessed
is incorrect. The precision is therefore 100\%
throughout, while the number of correct answers accepted
varies substantially across models and datasets. The confidence intervals show how much evidence supports
each precision estimate. On \textsc{FinanceBench},
Qwen2.5-7B and Llama-3.1-8B each have only ten correct
accepted answers, giving a 95\% interval of
[69.2\%, 100\%]. GPT-5.5 has 560 correct accepted
answers on \textsc{XBRLFiling}, with an interval of
[99.3\%, 100\%]. Thus, our observation is that identical observed precision
does not imply equally precise estimates. 

As shown in Figure~\ref{fig:combined-performance} (left), \textsc{VeriFin} rejects every
ground-truth-incorrect candidate produced by the evaluated generators
on \textsc{XBRLFiling}, yielding zero false accepts. Candidate claim
accuracy nevertheless varies widely, with cash-flow subtotals remaining particularly challenging (right).

\subsection{Exploratory Verifier-Guided Repair}
\label{sec:results_repair}

Table~\ref{tab:repair} presents an exploratory analysis of using
solver-derived diagnostics to guide repair attempts. On
\textsc{XBRLFiling}, naive-repair recovery is $6.5\%$, $26.1\%$, and
$23.1\%$ for Qwen3-30B, Fin-o1-8B, and Claude Haiku 4.5, respectively.
The corresponding verifier-guided pass@1 rates are $46.3\%$, $57.0\%$,
and $40.5\%$, while pass@3 reaches $47.7\%$, $69.9\%$, and $55.6\%$.
On FinanceBench, generic retries recover no cases in the scored subsets,
whereas pass@3 reaches $50.0\%$ for Qwen3-30B, $9.1\%$ for Fin-o1-8B,
and $20.0\%$ for Claude Haiku 4.5.

These results suggest that solver-derived feedback may help models
revise rejected answers and that additional guided attempts may provide
further recovery. However, because the conditions were evaluated on
separately scored archived subsets rather than a single matched set,
the results are descriptive rather than controlled comparisons.
Accordingly, this analysis provides an initial illustration of the
potential utility of verifier-guided repair.

\section{Error Analysis}

\paragraph{Abstentions.} Our manual review identified three recurring causes of abstention:
failure to establish a supported formula, resolve the required
evidence, or obtain a usable numerical claim. In several
evidence-related cases, the necessary values were already present
in the source. The difficulty was connecting those values to the
financial concepts required by the question, showing that evidence
availability alone does not ensure successful verification.

\paragraph{False Rejects.} Some false rejections arose from incorrect period selection,
misinterpreted financial quantities, and inconsistent sign
conventions. For example, a correct FY2025 answer was checked
against FY2026 facts, while another case treated net PP\&E
as gross PP\&E and applied an inappropriate depreciation
adjustment. These errors changed the financial problem encoded
in the constraints: a correct answer could then conflict with
the constructed check. They highlight the need to validate
the interpretation of financial inputs before symbolic checking.
We distinguished these construction errors from disagreements
between the evaluation and verification criteria.

\paragraph{Use of AI Assistants.}
Codex was used to assist with code implementation,
\LaTeX{} formatting, and editorial revisions of the manuscript.
The authors independently conducted all research, designed
the experiments, performed the analyses, and formulated
the scientific claims and conclusions.

\section{Conclusion}
\label{sec:conclusion}
We introduced \textsc{VeriFin}, a neurosymbolic framework
that verifies numerical financial claims against facts and
calculations established independently of the candidate answer.
Across six answer models on \textsc{XBRLFiling},
\textsc{FinanceBench}, and \textsc{TAT-DQA}, the reported
evaluations yield zero observed false accepts among claims
with known correctness. Baseline comparisons show that higher
precision can coincide with accepting more correct answers,
although the gains vary across models and datasets.
False rejections and abstentions highlight the importance of
correctly identifying financial concepts, selecting evidence,
and establishing the required calculation before symbolic
checking. These findings support neurosymbolic verification
for controlling false accepts in financial question answering.
Future work will focus on reducing abstentions through better
evidence resolution and financial interpretation while
maintaining high precision among accepted claims.

\bibliographystyle{ACM-Reference-Format}
\bibliography{bibliography}

@article{islam2023financebench,
  title={Financebench: A new benchmark for financial question answering},
  author={Islam, Pranab and Kannappan, Anand and Kiela, Douwe and Qian, Rebecca and Scherrer, Nino and Vidgen, Bertie},
  journal={arXiv preprint arXiv:2311.11944},
  year={2023}
}

@inproceedings{pan2023logic,
  title={Logic-lm: Empowering large language models with symbolic solvers for faithful logical reasoning},
  author={Pan, Liangming and Albalak, Alon and Wang, Xinyi and Wang, William},
  booktitle={Findings of the Association for Computational Linguistics: EMNLP 2023},
  pages={3806--3824},
  year={2023}
}

@inproceedings{wang2025finsage,
  title={Finsage: A multi-aspect rag system for financial filings question answering},
  author={Wang, Xinyu and Chi, Jijun and Tai, Zhenghan and Kwok, Tung Sum Thomas and He, Hailin and Li, Zhuhong and Hua, Yuchen and Li, Muzhi and Lu, Peng and Wang, Suyucheng and others},
  booktitle={Proceedings of the 34th ACM International Conference on Information and Knowledge Management},
  pages={6144--6152},
  year={2025}
}

@article{zheng2023judging,
  title={Judging llm-as-a-judge with mt-bench and chatbot arena},
  author={Zheng, Lianmin and Chiang, Wei-Lin and Sheng, Ying and Zhuang, Siyuan and Wu, Zhanghao and Zhuang, Yonghao and Lin, Zi and Li, Zhuohan and Li, Dacheng and Xing, Eric and others},
  journal={Advances in neural information processing systems},
  volume={36},
  pages={46595--46623},
  year={2023}
}

@article{chen2022program,
  title={Program of thoughts prompting: Disentangling computation from reasoning for numerical reasoning tasks},
  author={Chen, Wenhu and Ma, Xueguang and Wang, Xinyi and Cohen, William W},
  journal={arXiv preprint arXiv:2211.12588},
  year={2022}
}

@inproceedings{lai2025sec,
  title={Sec-qa: A systematic evaluation corpus for financial qa},
  author={Lai, Viet and Krumdick, Michael and Lovering, Charles and Reddy, Varshini and Schmidt, Craig and Tanner, Chris},
  booktitle={Proceedings of The 10th Workshop on Financial Technology and Natural Language Processing},
  pages={221--236},
  year={2025}
}

@article{guo2026finground,
  title={FinGround: Detecting and Grounding Financial Hallucinations via Atomic Claim Verification},
  author={Guo, Dongxin and Wu, Jikun and Yiu, Siu Ming},
  journal={arXiv preprint arXiv:2604.23588},
  year={2026}
}

@inproceedings{de2008z3,
  title={Z3: An efficient SMT solver},
  author={De Moura, Leonardo and Bj{\o}rner, Nikolaj},
  booktitle={International conference on Tools and Algorithms for the Construction and Analysis of Systems},
  pages={337--340},
  year={2008},
  organization={Springer}
}

@inproceedings{han2024xbrl,
  title={Xbrl agent: Leveraging large language models for financial report analysis},
  author={Han, Shijie and Kang, Haoqiang and Jin, Bo and Liu, Xiao-Yang and Yang, Steve Y},
  booktitle={Proceedings of the 5th ACM International Conference on AI in Finance},
  pages={856--864},
  year={2024}
}

@article{panda2026finverbench,
  title={FinVerBench: Benchmark Validity and Calibration in Large Language Model Financial Statement Verification},
  author={Panda, Silu},
  journal={arXiv preprint arXiv:2605.29586},
  year={2026}
}

@article{wang2026auditflow,
  title={AUDITFLOW: Executable Symbolic Environments for Structured Financial Reporting Verification},
  author={Wang, Yan and Ai, Xuguang and Patel, Jaisal and Peng, Xueqing and Mo, Fengran and Cao, Yupeng and Li, Haohang and Cao, Mingyu and Qian, Lingfei and Guti{\'e}rrez-Basulto, V{\'\i}ctor},
  journal={arXiv preprint arXiv:2606.03031},
  year={2026}
}

@article{jiang2026fin,
  title={Fin-rate: A real-world financial analytics and tracking evaluation benchmark for llms on sec filings},
  author={Jiang, Yidong and Chen, Junrong and Makri, Eftychia and Chen, Jialin and Li, Peiwen and Maatouk, Ali and Tassiulas, Leandros and Brenner, Eliot and Xiang, Bing and Ying, Rex},
  journal={arXiv preprint arXiv:2602.07294},
  year={2026}
}

@article{akinfaderin2025verafi,
  title={VERAFI: Verified Agentic Financial Intelligence through Neurosymbolic Policy Generation},
  author={Akinfaderin, Adewale and Subramanian, Shreyas},
  journal={arXiv preprint arXiv:2512.14744},
  year={2025}
}

@inproceedings{olausson2023linc,
  title={LINC: A neurosymbolic approach for logical reasoning by combining language models with first-order logic provers},
  author={Olausson, Theo and Gu, Alex and Lipkin, Ben and Zhang, Cedegao and Solar-Lezama, Armando and Tenenbaum, Joshua and Levy, Roger},
  booktitle={Proceedings of the 2023 Conference on Empirical Methods in Natural Language Processing},
  pages={5153--5176},
  year={2023}
}

@article{ye2023satlm,
  title={Satlm: Satisfiability-aided language models using declarative prompting},
  author={Ye, Xi and Chen, Qiaochu and Dillig, Isil and Durrett, Greg},
  journal={Advances in Neural Information Processing Systems},
  volume={36},
  pages={45548--45580},
  year={2023}
}

@article{bayless2025neurosymbolic,
  title={A neurosymbolic approach to natural language formalization and verification},
  author={Bayless, Sam and Buliani, Stefano and Cassel, Darion and Cook, Byron and Clough, Duncan and Delmas, R{\'e}mi and Diallo, Nafi and Erata, Ferhat and Feng, Nick and Giannakopoulou, Dimitra and others},
  journal={arXiv preprint arXiv:2511.09008},
  year={2025}
}

@inproceedings{lin2022truthfulqa,
  title={Truthfulqa: Measuring how models mimic human falsehoods},
  author={Lin, Stephanie and Hilton, Jacob and Evans, Owain},
  booktitle={Proceedings of the 60th annual meeting of the association for computational linguistics (volume 1: long papers)},
  pages={3214--3252},
  year={2022}
}

@inproceedings{min2023factscore,
  title={FActScore: Fine-grained atomic evaluation of factual precision in long form text generation},
  author={Min, Sewon and Krishna, Kalpesh and Lyu, Xinxi and Lewis, Mike and Yih, Wen-tau and Koh, Pang and Iyyer, Mohit and Zettlemoyer, Luke and Hajishirzi, Hannaneh},
  booktitle={Proceedings of the 2023 conference on empirical methods in natural language processing},
  pages={12076--12100},
  year={2023}
}

@inproceedings{zhao2024findver,
  title={FinDVer: Explainable claim verification over long and hybrid-content financial documents},
  author={Zhao, Yilun and Long, Yitao and Jiang, Tintin and Wang, Chengye and Chen, Weiyuan and Liu, Hongjun and Tang, Xiangru and Zhang, Yiming and Zhao, Chen and Cohan, Arman},
  booktitle={Proceedings of the 2024 Conference on Empirical Methods in Natural Language Processing},
  pages={14739--14752},
  year={2024}
}

@inproceedings{chungkham2025think,
  title={Think Right, Not More: Test-Time Scaling for Numerical Claim Verification},
  author={Chungkham, Primakov and Viswanathan, Venktesh and Setty, Vinay and Anand, Avishek},
  booktitle={Empirical Methods in Natural Language Processing (EMNLP) 2025},
  pages={24345--24363},
  year={2025},
  organization={Association for Computational Linguistics}
}

@inproceedings{zhu2022towards,
  title={Towards complex document understanding by discrete reasoning},
  author={Zhu, Fengbin and Lei, Wenqiang and Feng, Fuli and Wang, Chao and Zhang, Haozhou and Chua, Tat-Seng},
  booktitle={Proceedings of the 30th ACM International Conference on Multimedia},
  pages={4857--4866},
  year={2022}
}

\end{document}